\documentclass[12pt]{article}
\usepackage{cite,epsfig}
\usepackage{revsymb,epsfig}
\def\BA{\begin{eqnarray}} \def\BE{\begin{equation}}
\def\EA{\end{eqnarray}} \def\EE{\end{equation}} 
 
\def\gtsim{\lower-0.45ex\hbox{$>$}\kern-0.77em\lower0.55ex\hbox{$\sim$}}
\def\ltsim{\lower-0.45ex\hbox{$<$}\kern-0.77em\lower0.55ex\hbox{$\sim$}}
\begin{document}
\title{Vacuum Expectation Value of a Wegner-Wilson Loop Near the
Light-Cone} \author{Hans
J.~Pirner$^{ab}$\thanks{pir@tphys.uni-heidelberg.de}\ ,\
N. Nurpeissov$^{a}$ \\ {}\\ ${}^a$Institut f\"ur Theoretische Physik
der Universit\"at Heidelberg, Germany\\ ${}^b$Max-Planck-Institut
f\"ur Kernphysik Heidelberg, Germany} \maketitle
\begin{abstract}
Vacuum expectation values  for one Wegner-Wilson loop  representing a
moving quark-antiquark pair are calculated in four-dimensional
Euclidean and Minkowski space-time.  The calculation uses gluon field
strength correlators with perturbative gluon exchange and
non-perturbative correlations  from the stochastic vacuum model. The
expectation value of a Wegner-Wilson loop forming a hyperbolic angle
in Minkowski space-time is connected by an analytical continuation to
the expectation value of the  Wegner-Wilson loop  in Euclidean
four-space. The obtained result shows how confinement enters into the
light-cone Hamiltonian for valence quarks independently of the chosen
model.
\end{abstract}

\newpage
\section{Introduction}
One of the challenges in quantum chromodynamics (QCD) is the
relativistic bound state problem. In the light-cone Hamiltonian
approach \cite{Brodsky:1997de} light-cone wave functions  can be
constructed in a boost invariant way.  It is necessary to have
reliable light-cone wave functions  if one wants to calculate high
energy scattering, especially exclusive reactions. Many
parametrizations assume  separability of the dependence on the
longitudinal momentum fraction and transverse momentum which is very
unlikely since the two momenta are coupled in the kinetic energy
operator.  Various approaches have been tried to compute such wave
functions.  One can use the usual equal time Hamiltonian
\cite{Simula:2002vm} and transform the resulting wave functions into
light-cone form with the help of  kinematical on-shell equations. The
light-cone Hamiltonian in a string picture is formulated in
ref. \cite{Morgunov:vy}. More ambitious is the construction of an
effective  Hamiltonian including the gauge degrees of freedom
explicitly  and then solving the bound state problem. For mesons this
approach \cite{Burkardt:2001jg,Dalley:2002nj} still needs many
parameters to be fixed.  Attempts have been made to solve the valence
quark wave function for mesons  in a simple Hamiltonian with a
two-body potential \cite{Frederico:2002vs}.

A necessary input is an adequate potential for the light-cone
Hamiltonian. For the equal time Hamiltonian  and heavy quarks the
calculation of Wegner-Wilson loops gives the form of the
non-perturbative potential for long distances. The   correlator model
\cite{Sho:1} allows to calculate  vacuum expectation values of
gauge invariant Wegner-Wilson loops using perturbative and
non-perturbative field strength correlation functions as input.  One
computes the loop expectation value $\left<W_r[C]\right>$ in terms of
a gauge invariant bilocal  gluon field strength correlator integrated
over minimal surfaces by using non-Abelian Stokes' theorem.  Then the
matrix cumulant expansion in the Gaussian approximation is applied.

The basic object of the correlator model is
the
gauge invariant bilocal gluon field strength correlator 
$F_{\mu\nu\rho\sigma}(X_1,X_2,O;C_{x_1o},C_{x_2o})$. The strings
$C_{x_1o},\,C_{x_2o}$ connect the coordinates $X_1,\,X_2$ in  the
correlation function  of the two-field strengths to a common
reference point $O$. We define
\begin{equation}
\frac{1}{4}\delta^{ab}F_{\mu\nu\rho\sigma}(X_1,X_2,O;C_{x_1o},C_{x_2o}):=
\left<\frac{g^2}{4\pi^2}\left[G^a_{\mu\nu}(O,X_1;C_{x_1o})
G^b_{\rho\sigma}(O,X_2;C_{x_2o})\right]\right>_G.
\label{eq_13}
\end{equation}
 The gluon field correlator has a perturbative ($P$) and a non-perturbative
 ($NP$) component.  The stochastic vacuum model is used for the
non-perturbative low frequency background field and  the  perturbative
gluon exchange for the additional high frequency contributions. 
The most general form of the correlator respecting translational,
Lorentz and parity invariance reads in Euclidean space \cite{Sho:1}%
\begin{eqnarray}
F^{NP}_{\mu\nu\rho\sigma}(Z)& = & F^{NP\,c}_{\mu\nu\rho\sigma}(Z)+
F^{NP\,nc}_{\mu\nu\rho\sigma}(Z)\nonumber\\ & = & \frac{1}{3(N_c^2-1)}
G_2 \Biggl\{\kappa(\delta_{\mu\rho}\delta_{\nu\sigma}-
\delta_{\mu\sigma}\delta_{\nu\rho})\,D(Z^2)
\label{eq_1_20}\\
&   & +(1-\kappa)\frac{1}{2}\left[\frac{\partial}{\partial Z_\nu}
      (Z_\sigma \delta_{\mu\rho}-Z_\rho\delta_{\mu\sigma})+
      \frac{\partial}{\partial Z_\mu}(Z_\rho\delta_{\nu\sigma}-
      Z_\sigma\delta_{\nu\rho})\right]D_1(Z^2)\Biggr\}\nonumber
\end{eqnarray}
with
$G_2=\left<\frac{g^2}{4\pi^2}G^a_{\mu\nu}(O)G^a_{\mu\nu}(O)\right>$
as the gluon condensate. The term proportional to $\kappa$ is the
non-Abelian confining part $F^{NP\,c}$ of the correlator, in contrast,
the tensor structure $F^{NP\,nc}_{\mu\nu\rho\sigma}$ is characteristic
for Abelian  gauge theories and does not lead  to confinement.  The
correlation functions are a simple exponential of range a: \BE
D(Z^2)=D_1(Z^2)=e^{-|Z|/a}.
\label{eq_exp_cor}
\EE

The
calculation of a Wegner-Wilson loop along the imaginary time
directions gives the heavy quark-antiquark potential with
color-Coulomb behavior for small  and confining linear rise for large
source separations \cite{Sho:1}.
Since the computation of the VEV for one Wegner-Wilson loop can be
done completely analytically,  also other  orientations of the loop
can be chosen, e.g. a loop where the quark-antiquark  pair moves along
the z-direction.  By transforming to Minkowski space-time the
dependence of the interaction potential on longitudinal and transverse
separation of the pair can be obtained this way.  In section 2 we
describe the  calculation in Euclidean space-time, in section 3 in
Minkowski space time and in section 4 we derive the potential in a
light-cone Hamiltonian for valence quarks.

\section{Vacuum expectation value for a tilted
Wegner-Wilson loop in Euclidean space-time}

\begin{figure}[ht]
\includegraphics{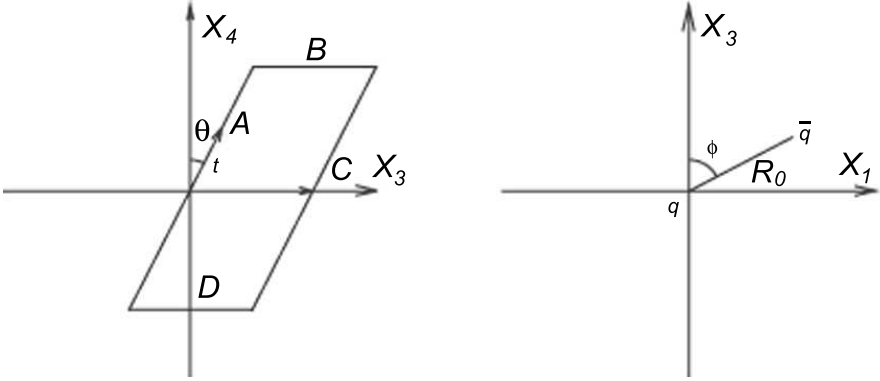}
\caption{Configuration of the Wegner-Wilson loop in Euclidean
space-time.}
\label{loop_eu}
\end{figure}

The vacuum expectation value (VEV) of a tilted
Wegner-Wilson loop represents a moving quark-antiquark pair  
\begin{equation}
W[C]={Tr}\mathcal{P} exp {\left(-ig
\oint\limits_{C}dZ_{\mu}G^a_{\mu}(Z) t^a\right)}.
\label{eq_WWL}
\end{equation}
The  group generators $t^a$ are in the
fundamental representation of $SU(3)$, $g$ is the strong coupling constant
and $\cal{P}$ symbolizes path ordering
of the closed path $C$ in  space-time.
The loop $C  $ with spatial extension $R_0$ and temporal extension $
T$   has the following parametrization in  four-dimensional Euclidean space-time
(Fig. \ref{loop_eu})

\begin{displaymath}
C=C_A\bigcup C_B\bigcup C_C\bigcup C_D,
\end{displaymath}
where%
\begin{eqnarray}
C_A & = & \left\{u t_{\mu},  \qquad u\in [-T/2,T/2] \right\}
\nonumber\\ C_B & = & \left\{ T/2 t_{\mu}+v R_0 r_{\mu},  \qquad v\in
[0,1]\right\} \nonumber\\ C_C & = & \left\{-u t_{\mu}+ R_0 r_{\mu},
\qquad u\in [-T/2,T/2] \right\}
\label{eq_1_21}\\
C_D & = & \left\{-T/2 t_{\mu}+ (1-v) R_0 r_{\mu},  \qquad  v\in
[0,1]\right\} \nonumber
\end{eqnarray}
and the parametrization of the surface%
\begin{equation}
X_\mu=u t_\mu \,+\,v R_0 r_\mu,\quad u\in[-T/2,T/2]; \:v\in[0,1],
\end{equation}
where%
\begin{eqnarray}
r       & = & \left(\begin{array}{c} \sin{\phi}  \\ 0 \\ \cos{\phi} \\
0 \\
\end{array}\right) \nonumber\\
t       & = & \left(\begin{array}{c} 0 \\ 0 \\ \sin{\theta} \\
\cos{\theta} \\
\end{array}\right).
\end{eqnarray}

The expression for the Wegner-Wilson loop simplifies with the help of
the Casimir operator in the fundamental representation $C_2(3)=t^2=4/3$
\begin{equation}
\left<W[C]\right>_G\,=\,\exp{\left[-\frac{C_2(3)}{2}\chi_{ss}\right]},
\label{eq_15}
\end{equation}
where $\chi_{ss}$ is the double area integral of the
correlation function over the surface
\begin{equation}
\chi_{ss}\,:=\,\frac{\pi^2}{4} \int\limits_{S}
d\sigma_{\mu\nu}(X_1)\int\limits_{S} d\sigma_{\rho\sigma}(X_2)
\,F_{\mu\nu\rho\sigma}(X_1,X_2,O;C_{x_1o},C_{x_2o}).
\label{eq_chi_euc}
\end{equation}
The lengthy calculation of $\chi_{ss}$ is standard and follows the
lines of reference \cite {Sho:1}, we will give here only these
parts which are relevant to understand the calculation of the
tilted loops.
One gets for the
non-perturbative confinement component
\begin{eqnarray}
\chi^{NP\,c}_{ss}&=&\frac{\pi^2 G_2 \kappa}{3(N_c^2-1)}\int\limits_0^1
dv_1 \int\limits_0^1 dv_2 \int\limits_{-T/2}^{T/2} du_1
\int\limits_{-T/2}^{T/2} du_2  \left[t^2\cdot r^2-(t\cdot
r)^2\right]\,D(Z^2)\\ &=&\frac{\pi^2 G_2 \kappa}{3(N_c^2-1)} R_0^2
(1-\cos^2{\phi} \sin^2{\theta}) \int\limits_0^1 dv_1 \int\limits_0^1
dv_2 \int\limits_{-T/2}^{T/2} du_1 \int\limits_{-T/2}^{T/2} du_2
\,D(Z^2).
\end{eqnarray}
Correlated points on the surface have the distance
$Z=(u_1-u_2)t + (v_1-v_2)R_0 r$.
The geometry of the loop orientation enters via the factor  $\alpha$
\BE \alpha^2=1-\cos^2{\phi}\sin^2{\theta}.  \EE

The confining $\chi_{ss}$ has the
following final form:
\begin{eqnarray}
\chi^{NP\,c}_{ss}       =  \lim\limits_{T\rightarrow\infty}{}
        \frac{2\pi^2 G_2 \kappa T}{3(N_c^2-1)}
        \int\limits_0^{R_0\alpha} d\rho (R_0\alpha-\rho) \cdot
        2\rho\,K_1\left(\frac{\rho}{a}\right).
\label{eq_1_19}
\end{eqnarray}
At large distances $R_0 \alpha>> 2 a $ 
one recognizes that the confining interaction leads to a VEV of the
tilted Wilson loop which is consistent with the area law 
$R_0$ 

\BA 
<W[C]>&=&e^{-\sigma R_0 \alpha T}\\
\sigma&=&\frac{\pi^3 G_2a^2 \kappa}{18}, \EA where $\sigma $ is the
string tension \cite {Sho:1} and the area is obtained from \BA Area&=&
T R_0 \int_{-1/2}^{1/2}du \int_0^1 dv
\sqrt{\left(\frac{dX_{\mu}}{du}\right)^2\left(\frac{dX_{\mu}}{dv}\right)^2
-\left(\frac{dX_{\mu}}{du}\frac{dX_{\mu}}{dv}\right)^2}\\ &=&T R_0
\alpha.  
\EA

The non-confining $\chi_{ss}$ functions give the short range
attractive quark-antiquark interaction from massive correlator and
gluon exchange 
\BA
\chi^{NP\,nc}_{ss}&=&-\lim\limits_{T\rightarrow\infty}{} \frac{ 2
\pi^2 G_2 (1-\kappa) a T}{3(N_c^2-1)} \cdot R_0^2\alpha^2 \cdot
K_2\left(\frac{R_0\alpha}{a}\right)\\ \chi^P_{ss} &=&
-\lim\limits_{T\rightarrow\infty}{} \frac{ 2 g^2 T \exp{\left(-m_G
R_0\alpha\right)}} {4\pi R_0\alpha}\ .
\label{eq_concl_3}
\EA
In the limit of straight loops  ($\theta=0$) all results agree with
previous calculations \cite {Sho:1}.

\section{Vacuum expectation value for one 
Wegner-Wilson loop in Minkowskian space-time near the light-cone}

In this section the vacuum expectation value (VEV) of one
Wegner-Wilson  loop near the light-cone is computed 
in Minkowskian space-time.  As before we use the correlator
model for the non-perturbative low  frequency background
field and perturbative gluon exchange for  the additional
high frequency  contribution. The path of the color dipole in 
four-dimensional  Euclidean space-time is represented  by a light-like QCD
Wegner-Wilson loop: To accomplish the transition  from Euclidean to
Minkowski space-time we have to make the following replacements:
\begin{eqnarray}
X_4  & \longrightarrow & i x^0 \nonumber\\ X_i & \longrightarrow & x^i.
\nonumber\\
\end{eqnarray}
Here $X=(\vec{X},X_4)$ is the Euclidean space-time point and
$x=(x^0,\vec{x})$ the Minkowskian vector. Of course, also the
Euclidean correlation functions of the field strengths have to be
changed to the Minkowskian correlation functions \cite {steffen}.

\begin{figure}[ht]
\centering \includegraphics{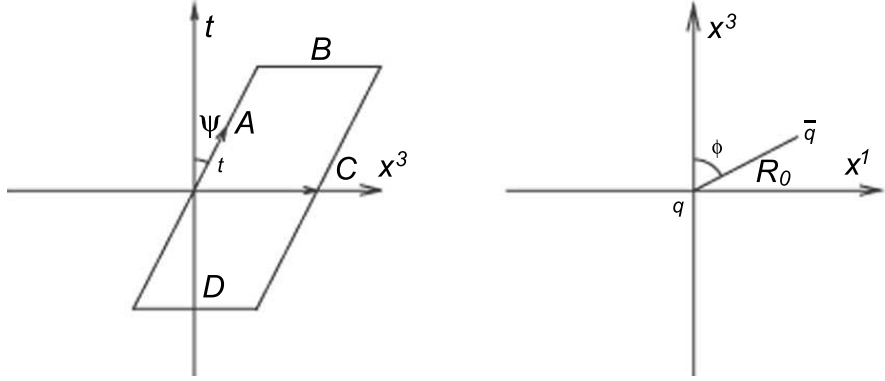}
\caption{Configuration of the Wegner-Wilson loop in Minkowski
space-time.}
\label{loop_mink}
\end{figure}

The loop $C$ with spatial extent $R_0$ and temporal extension $T$
is now placed in four-dimensional Minkowski space-time.  The
quark-antiquark pair is moving with velocity $\beta$  \BE
\beta=\frac{\sinh(\psi)}{\cosh(\psi)} \EE and the hyperbolic angle
$\psi$ defines the boost  (Fig. \ref{loop_mink}).

The loop has the same parametrization as before but with Minkowskian
vectors $r^{\mu}$ and $t^{\mu}$

\begin{eqnarray}
r^{\mu}       & = & \left(\begin{array}{c} 0  \\ \sin{\phi} \\ 0 \\
\cos{\phi} \\
\end{array}\right) \nonumber\\
t^{\mu}       & = & \left(\begin{array}{c} \cosh{\psi} \\ 0 \\ 0 \\
\sinh{\psi} \\
\end{array}\right).
\end{eqnarray}


For the
Wegner-Wilson loop in Minkowski space-time we define $\chi_{ss}$  in
the same way as in ref. \cite{steffen} 

\BE \left<W[C]\right>_G =
\exp{\left[-i\,\frac{C_2(3)}{2}\chi_{ss}\right]}.  \EE 

The phase
factor $\chi_{ss}$ is given as the double area integral over the
surface 

\BE \chi_{ss}\,:=\,-i\,\frac{\pi^2}{4} \int\limits_{S}
d\sigma^{\mu\nu}(x_1)\int\limits_{S} d\sigma^{\rho\sigma}(x_2)
\,F_{\mu\nu\rho\sigma}(x_1,x_2,O;C_{x_1o},C_{x_2o}).  \EE 

The
expression for the Wegner-Wilson loop is the same as before
eq. (\ref{eq_15}) and eq. (\ref{eq_chi_euc}). We use  the above  convention
in order to be consistent with the notation of ref. \cite{steffen}.  In
the course of the calculation we find%

\begin{equation}
\chi^{NP\,c}_{ss}=-\frac{\pi^2 G_2 \kappa}{12(N_c^2-1)}\int\limits_S
d\sigma^{\mu\nu} (x_1) \int\limits_S d\sigma^{\rho\sigma} (x_2)
\,iD(z^2/a^2)  \left\{g_{\mu\rho} g_{\nu\sigma}- g_{\mu\sigma}
g_{\nu\rho}\right\},
\end{equation}

where the changed metric tensor in Minkowski space comes from
replacing the Euclidean metric tensor by the Minkowski tensor in the
correlation function.  Using the Minkowskian vectors $r,t$ with the
following properties

\begin{eqnarray*}
t^2             & = & \cosh^2 {\psi}-\sinh^2 {\psi}=1\\ r^2
& = & r_\mu r^\mu = -1\\ (t\cdot r)^2    & = & (t_\mu r^\mu)\cdot
(t^\nu r_\nu)= \cos^2{\phi} \sinh^2{\psi}\\ t^2 \cdot r^2 - (t\cdot
r)^2 & = & -(1+\cos^2{\phi} \sinh^2{\psi})
\end{eqnarray*}

one obtains

\begin{displaymath}
\chi^{NP\,c}_{ss}=\lim\limits_{T\rightarrow\infty}{} \frac{2\pi^2 G_2
\kappa T}{3(N_c^2-1)}  \int\limits_0^{R_0\alpha_M} d\rho
(R_0\alpha_M-\rho)\, iD^{(3)}(\rho^2).
\end{displaymath}

The three-dimensional  correlation function $D^{(3)}(\rho^2)$ in
Minkowski space  is obtained by analytical continuation of the
Euclidean function.  Since the argument of the function is given by
the magnitude of a three-vector the analytic continuation leads to the
same  modified Bessel function which was obtained in Euclidean space,
cf. also Appendix B of ref. \cite{steffen}. The resulting  $\chi_{ss}$
is real.  The geometry enters via the factor 

\BE
\alpha_M^2=1+\cos^2{\phi}\sinh^2{\psi} \EE

which is consistent with the  analytical continuation of the Euclidean
expression $\alpha=1-\cos^2{\phi}\sin^2{\theta}$ into Minkowski space
by transforming the angle $\theta \rightarrow i \psi$.  This
analytical continuation is similar to the analytical continuation used
in high energy scattering
\cite{Meggiolaro:1996hf,Meggiolaro:2001xk,Hebecker:1999pb} where the
angle between two Wilson loops transforms in  the same way.

The different contributions to $\chi_{ss}$ read: \BA \chi^{NP\,c}_{ss}
& = & \lim\limits_{T\rightarrow\infty}{} \frac{2\pi^2 G_2 \kappa
T}{3(N_c^2-1)} \int\limits_0^{R_0\alpha_M} d\rho (R_0\alpha_M-\rho)
\cdot 2\rho\,K_1\left(\frac{\rho}{a}\right)\nonumber\\
\chi^{NP\,nc}_{ss}&=&-\lim\limits_{T\rightarrow\infty}{} \frac{ 2
\pi^2 G_2 (1-\kappa) a T}{3(N_c^2-1)} \cdot R_0^2\alpha_M^2 \cdot
K_2\left(\frac{R_0\alpha_M}{a}\right)\\ \chi^P_{ss} &=&
-\lim\limits_{T\rightarrow\infty}{} \frac{ 2 g^2 T \exp{\left(-m_G
R_0\alpha_M\right)}} {4\pi R_0\alpha_M}.  \EA

We remark that the same calculation with a time-like vector $t \cdot
t=0$ and a quark separation $r^{\mu}$ oriented orthogonally to $t$,
i.e. $r\cdot t=0$ gives the well-known result that the expectation
value of the loop equals unity as has been shown in the original
references on  high energy scattering. In the formulation of high
energy scattering \cite {steffen} with the correlation function summed
to  all orders in the S-matrix the respective phase factors from
single  loops totally cancel.  So the here obtained expectation value
does not change  the results of the previous calculations even if one
goes away from the light-cone.

\section{Using the VEV of the light-like Wilson loop to
derive the quark-antiquark potential in the light-cone Hamiltonian}

The exponent giving
the expectation value of the Wilson loop acquires a new meaning 
for a tilted loop.
In order to interprete the result of the preceding section one must
define the four-velocity of the particles described by the tilted loop

\BE u_{\mu}= (\gamma,0_\bot,\gamma \beta).  \EE 

With the help of the four-velocity we can rewrite the loop as:

\BE 
e^{- ig \int d\tau A^{\mu}u_{\mu}} =e^{- ig \int d\tau  ( \gamma A^0-
\gamma \beta A^3)}.
\EE

The line integral of the gauge potential  acts as a phase factor  
on a Dirac wave function $\psi$ which splits
up into a leading dynamical component $\psi_+$ and a
dependent component $\psi_-$. For very fast quarks the mass term and
transverse momenta  are negligible compared with the energy and
longitudinal momentum. In this eikonal approximation the Dirac equation of 
the leading component decouples from the small component:

\BA i \partial_- \psi_+ &=&  P^-_{pot}(A^-)  \psi_+\\
                        &=&  g A^- \psi_+. 
\EA

With $ \beta \approx 1$ the phase factor in the tilted Wilson loop
integrates $A^-$ and leads to a VEV for the loop containing 
$P^-_{pot}=\frac{1}{\sqrt{2}}(P^0-P^3)|_{pot}$

\BE <W_r[C]>=e^{-i \gamma(P^0-P^3)|_{pot} T}.  \EE 

The light-cone potential energy arising from
the confining part of the correlation function  has a term of order 
$O(\frac{1}{P^+})=O(\frac{1}{\gamma})$, where $P^+=\frac{1}{\sqrt{2}}(P^0+P^3)$ is 
the light-cone momentum

\BE P^-_{pot}=
\frac{1}{\sqrt{2}}\left(\sigma R_0 \sqrt{\cos(\phi)^2+\sin(\phi)^2/
\gamma^2}\right).  \EE

Terms involving transverse momenta and masses of the
same order $O(\frac{1}{P^+})$ are not included in the loop
as it has been calculated. 
Two of these terms give the standard kinetic energy
term of  free particles, which 
contributes to the total light-cone energy. 
Terms with spin cannot be 
obtained from the straight Wilson loop and are not discussed here.
We introduce the relative $+$ momentum $k^+$ and
transverse momentum $k_{\bot}$ for the quarks with  mass $\mu$.  By
adding the above ``potential'' term to the kinetic term of relative
motion  of the  two particles we complete our
approximate derivation of the light-cone energy $P^-$

\BE P^-= \frac{(\mu^2+ k_{\bot}^2)P^+}{2 (1/4 P^{+2}-k^{+2})}+
\frac{1}{\sqrt{2}}\sigma \sqrt{x_3^2+x_{\bot}^2/\gamma^2}.  \EE 

To derive the light-cone Hamiltonian  we multiply $P^-$ with the plus
component of the light-cone momentum
$P^+=\frac{1}{\sqrt{2}}(P^0+P^3)$ and use that $P^+/M=\sqrt {2}
\gamma$
 to eliminate the boost variable from the Hamiltonian.  
Further we
follow the notation of reference \cite{Bardeen:1975gx} and introduce
the fraction $\xi=k^+/P^+$ with $|\xi| < 1/2$ and its conjugate  the
scaled longitudinal space coordinate $\sqrt{2} \rho= P^+ x_3$ as
dynamical variables.  For our configuration the relative time of the
quark and antiquark is zero

\BE M^2_c=2 P^+ P^-=  \frac {(\mu^2+ k_{\bot}^2)} {1/4-\xi^2} +  2
\sigma\sqrt{\rho^2+ M^2_c x_{\bot}^2}.  \EE

We have obtained the light-cone Hamiltonian $M^2_c$ from the
confining interaction in a Lorentz invariant
manner, because the variables $\xi,\rho,k_{\bot}$ and $x_{\bot}$ are
invariant under boosts.  The valence quark light-cone Hamiltonian has
a simple  confining potential. The magnitude of the confining
potential is set by the string tension $\sigma$. The effective
``distance'' of the quarks is given by  scale-free light-cone
longitudinal distance and the transverse distance multiplied by the
bound state mass.  The above equation agrees in the limit of
one-dimensional motion with the equation for the yo-yo string derived in
ref. \cite{Bardeen:1975gx}. If there is only transverse motion $(
\xi=0)$, the confinement has the usual form which is seen by using $M
\approx 2 \mu$. To solve the $M^2_c$ operator one can go over to
$M^2_s$. Minimizing  $M^2_s$ with respect to 
$s$ one can  replace the 
square root operator. Final self consistency must be reached with a guessed mass
eigenvalue $M_0$

\BE  M^2_s =  \frac {(\mu^2+ k_{\bot}^2)} {1/4-\xi^2} +
\frac{1}{2}\left( 4 \sigma^2 \frac {\rho^2+ M_0^2 x_{\bot}^2}{s} +s
\right). \EE

The other non-confining potentials from the abelian correlator and the
perturbative gluon exchange can be worked out similarly and one gets
for the complete valence Hamiltonian

\BE M^2 = \frac{(\mu^2+
k_{\bot}^2)} {1/4-\xi^2}  + 2 \sigma r  -4/3 \left( \frac{2 g^2 M^2
e^{-\frac{m_G r}{M}}}{4 \pi r}\right) -2
\frac{\sigma(1-\kappa)r^2}{\kappa M a \pi}K_2(\frac{r}{M a})  \EE 

with the dimensionless variable

\BE r=\sqrt{\rho^2+ M^2 x_{\bot}^2}. \EE

The relative weight of non-perturbative non-abelian and abelian
contribution is fixed by $\kappa=0.7$ in the parametrization of the
correlation function \cite{steffen}. We have used $\sigma$ to
parametrize also the abelian non-confining potential, instead of giving
the full expression with the gluon condensate. Of course, the abelian
part of the potential does not confine.  The best way to find the 
two-body wave function is to keep  $\xi$ in the momentum representation
and the transverse direction in configuration space $x_{\bot}$.  It is
to be expected that in this approximation the pion mass is not
described  correctly. Firstly, the spin structure of the meson is not
reflected in the spin independent expression above. Secondly, one
expects quark self energy corrections \cite{Simonov:2001iv}. The
typical mass scale of a meson estimated from a trial solution with a
wave function $\psi( \xi, x_{\bot}) = A \,cos( \xi \pi)
e^{-x_{\bot}^2/x_0^2}$ comes out to be 1 GeV.  The problem of the pion
has to be addressed separately, since on the light-cone  the mechanism
of chiral symmetry breaking needs special care and is of particular
interest. In the approach above confinement plays an important role in
contrast to Nambu-Jona-Lasinio effective models which give an adequate
description of spontaneous chiral symmetry breaking but do not include
confinement.

The confining interaction in the light-cone Hamiltonian was derived in the
specific model of the stochastic vacuum. But it also can be inferred
from the simple Lorentz transformation properties of the phase in the
Wilson loop and a lattice determination of the tilted Wilson
expectation values. In this respect the final Hamiltonian is model
independent.

The inclusion of confining forces in the initial and final state wave
functions can put all scattering cross sections  calculated with the
stochastic vacuum model on a much safer base  since wave functions
and cross section are derived consistently.  For low $Q^2$ the long
distance part of the photon wave function matters strongly and
confinement is important cf. \cite{Dosch:1997nw}.  Especially the
diffractive cross section has a sizeable  contribution from large
dipole sizes and a correct  behaviour can only be expected when the
problem of the large dipole wave function is treated adequately.  A
very useful extension of the above calculation is the coupling of the
initial $q \bar q$ state to higher Fock states $q \bar q g$ 
with gluons or fragmentation of the original  $q \bar q$ state 
into $q \bar q q \bar q$ states 
which can also  be estimated  from Wilson loops near the light-cone
 in Minkowski space. On the problem of fragmentation
the Lund model \cite{Andersson:tv} has been very successful and it is interesting
to see
how the above model calculation fares in comparison.

\vspace{1.0cm}

{\bf Acknowledgments}: We thank A. Hebecker  for bringing this problem
again to our attention and our colleague H.G. Dosch for a helpful
discussion.


\end{document}